\documentclass[aps,twocolumn]{revtex4-2}
\usepackage{graphicx}
\usepackage{color}
\usepackage{ amssymb }
\newcommand{\Prob}{\mathrm{P}}
\newcommand{\B}{{\cal B}}
\newcommand{\best}{\mathrm{best}}
\newcommand{\ccs}{\mathrm{ccs}}
\newcommand{\dep}{\mathrm{dep}}
\newcommand{\pool}{\mathrm{pool}}
\newcommand{\init}{\mathrm{init}}
\newcommand{\final}{\mathrm{final}}
\newcommand{\pos}{\mathrm{poss}}
\newcommand{\info}{\mathrm{Information}}

\newcommand{\eea}{\end{eqnarray}}
\newcommand{\bea}{\begin{eqnarray}}
\newcommand{\ee}{\end{equation}}
\newcommand{\be}{\begin{equation}}
\newcommand{\varT}{{\cal T}}

\newcommand{\I}{\tilde{I}}
\newcommand{\tH}{\tilde{H}}
\newcommand{\unique}{{\rm unq}\,}

\newcommand{\shared}{{\rm red}\,}
\newcommand{\synergy}{{\rm syn}\,}
\newcommand{\syn}[2]{\synergy #1 \& #2}
\newcommand{\unq}[2]{\unique #1 \backslash #2}
\newcommand{\red}[2]{\shared #1 \& #2}
\newcommand{\setn}[1]{\{#1\}}
\newcommand{\bhat}{Bhattacharyya\,}
\begin{document}
\title{Pooling Probability Distributions and the Partial Information Decomposition}
\author{S.J. van Enk}
\affiliation{Department of Physics,\\
University of Oregon, Eugene, OR 97403}
\begin{abstract}
Notwithstanding various attempts to construct a Partial Information Decomposition (PID) for multiple variables by defining synergistic, redundant, and unique information,
there is no consensus on how one ought to precisely define either of these quantities. One aim here is to illustrate how that ambiguity---or, more positively, freedom of choice---may arise. Using the basic idea that information equals the average reduction in uncertainty when going from an initial to a final probability distribution, synergistic information will likewise be defined as a difference between two entropies.
One term is uncontroversial and characterizes ``the whole'' information that source variables carry jointly about a target variable $\varT$. The other term then is meant to characterize the information carried by the ``sum of its parts.'' Here we interpret that concept as needing a suitable probability distribution aggregated (``pooled'') from multiple marginal distributions (the parts).  
	Ambiguity arises in the definition of the optimum way to pool two (or more) probability distributions.
	Independent of the exact definition of optimum pooling,
the concept of pooling leads to a lattice that differs from the  often-used redundancy-based lattice.
	One can associate not just a number (an average entropy) with each node of the lattice, but (pooled) probability distributions. As an example,
	one simple and reasonable approach to pooling is presented, which naturally gives rise to the 
overlap between different probability distributions as being a crucial quantity that characterizes both synergistic and unique information.
\end{abstract}
\maketitle
\section{Introduction}
Since the seminal work by Williams and Beer \cite{williams2010} on the partial information decomposition (PID) and their proposed definitions  for synergistic, redundant, and unique information, a lot of progress has been made to further clarify these notions. 
While the intuitive notions seem fairly clear at first sight, upon closer study there is some ambiguity left that has been surprisingly difficult to eliminate, witness the many valuable but different proposals for defining an explicit PID \cite{harder2013,bertschinger2014,griffith2014,ince2017,rauh2017secret,finn2018,james2018,james2018b,ay2019}. 
One point of the current work is to locate where that ambiguity may arise. The other is to question whether all terms in the decomposition correspond to information.

There are various reasons for desiring a PID. 
Ref.~\cite{james2017} gives a nice example of two probability distributions over three variables that, although produced by clearly distinct mechanisms, cannot be distinguished by just using the standard mutual informations between the different combinations of the three variables. But since those two distributions can be distinguished by using a PID, this demonstrates an immediate use of that decomposition as revealing different underlying mechanisms. In that role, a PID may help our understanding of complex networks  \cite{battiston2021,rosas2019}, such as neural networks, either involving actual neurons or artificial ones \cite{tax2017,wibral2017a,wibral2017b}.

Moreover, synergistic information may explain emergence \cite{varley2022,mediano2022} as implementing the adage that ``the whole is more than the sum of its parts.'' Perhaps a similar idea may even explain or define consciousness \cite{luppi2020}. 
For a perspective on the uses and prospects of PID, see Ref.~\cite{lizier2018}, and for 
discussions of important ideas that are made use of in the following, see Refs.~\cite{bertschinger2014,ince2017,kolchinsky2022}.
For earlier definitions of synergy and redundancy in the context of neural encoding outside the framework of PID, see Refs.~\cite{schneidman2003,latham2005}.

Here we propose a new perspective.
Just like Ref.~ \cite{gutknecht2021} went back to basics about part-whole relationships in order to get a fundamental idea of synergistic information, here we return to basics about information.
Within Shannon's theory \cite{shannon1948,cover1999}, information  is defined as the average reduction in uncertainty upon changing from an initial probability distribution to a final distribution.   Information is then always of the generic form
\be\label{basic}
\info=H_\init-H_\final, 
\ee
as a difference between two (averaged) entropies.
Moreover, information is always {\em about} something, and in our case the something will be represented by a target variable $\varT$. The uncertainty about $\varT$ taking on a particular value $t$ is quantified as $\log (1/\Prob(t))$, with the logarithm taken in base 2, and with $\Prob(t)$ the probability of finding the value $t$. This uncertainty is
averaged over all possible values $\varT$ can take  and over other initial and final variables, respectively, thus yielding expressions for the entropies $H_\init$ and $H_\final$ in Eq.~(\ref{basic}).
Typically, the relation between final and initial probability distributions is such that information is non-negative.

Using this basic idea, in order to define synergistic information, we will need, besides the joint distribution, a single distribution aggregated from multiple marginal distributions. 
To define such a distribution it seems easiest to use the language and techniques of ``pooling distributions'' for combining different expert opinions into one opinion (by compromise if not by consensus) \cite{lind1988,pool2000,carvalho2023,neyman2022}.
This language, although not necessary, is convenient for setting up notation, as explained in Sections \ref{experts}  and \ref{poolexp}.  In Section \ref{SUR} we define synergistic information in terms of a generic pooled probability distribution. We provide a side-by-side comparison of our PID with one concrete measure, taken from Ref.~\cite{bertschinger2014}, as both the analogies and the differences are clear and illuminating. 
Another comparison can be made by setting up an alternative PID by defining redundant information (as a difference between two averaged entropies) first.

The idea of pooling will help us define (in Section \ref{lattice}) a {\em lattice} \cite{lattice} underlying the structure of the PID for multiple variables, which differs from the redundancy-based lattice used in the original Williams-Beer work \cite{williams2010}.

Concrete ways of pooling probability distributions
are discussed in Section \ref{pool}. For one convenient and popular way of pooling, synergistic information and unique information acquire natural interpretations, as shown in Section \ref{simple}.
Examples illustrating the consequences for a PID based on pooling are given in Section \ref{Examples}.

\section{Setup and notation}
\subsection{Experts and their opinions}\label{experts}
We may use the following scenario to set terminology and notation.  Capital letters indicate variables, lower-case letters indicate their possible values. Entropies and information are expressed in units of bits.

Suppose we are interested in a variable $\varT$. We hire several experts (or agents), Alice, Bob, Charlie, \ldots to help us predict which of possible (discrete) values $t_1,t_2,\ldots$ $\varT$ may take on under specific conditions. More precisely, each expert $X$ is expected to report to us a probability distribution $\Prob_X(\varT)$.

Each expert has their own lab and each performs their own measurements of a single variable, which is indicated by the agent's initial $A,B,C,\ldots$. Each expert measures how their variable correlates with $\varT$. That is, each expert estimates a probability distribution $P_X(\varT):=P(\varT|X)$ for $X=A,B,C,\ldots.$ \footnote{For the goal of defining the PID it is assumed that these experimentally estimated conditional probability distributions are accurate to an arbitrary degree. Effects of statistical fluctuations in estimates of probability distributions for which the PID is to be determined seem not to have been considered.}.
Expert $X$'s uncertainty regarding $\varT$ is quantified by the Shannon entropy \cite{shannon1948} of the conditional distribution
\be
H_{\setn{X}}=H(\varT|X)=\sum_x\sum_t \Prob(x,t) \log(1/\Prob(t|x)).
\ee
The subscript uses set notation, this being convenient for the general definition of the PID. In this case, we have one set containing one expert, $X$. We may make this definition of uncertainty operational by imagining that we impose a fine on expert $X$ equal to 
\be\label{fine}
{\rm Fine}_X(t)=\log(1/\Prob_X(t)),
\ee
 if the value $t$ actually occurs. (The concept behind this operational definition goes by the name of ``scoring rules'' and one important feature of this specific rule is that the expert is forced to report their best (or ``true'') probability distribution if they want to minimize their fine \cite{scoring}.)

We assume here consistency among all experts, in that the unconditioned distribution for $\varT$ is identical for all agents. 
In other words, we assume here that there is a joint distribution
$\Prob(\varT,A,B,C\ldots)$ from which all other distributions can be obtained by marginalizing over some or all of the variables $A,B,C\ldots$. \footnote{This is a strong assumption, which would not apply to opinions held by meteorologists about the weather or by economists about inflation, unless they all used the same weather or inflation model, respectively. It does hold for the typical models for which the PID is meant to be used.}
In particular, we may define
\be
\Prob_{\varnothing}(t)=\sum_a \Prob_A(t|a)=
\sum_b\Prob_B(t|b)=\ldots
\ee 
with the empty set in the subscript indicating we need no expert opinions for this probability distribution.
The uncertainty we have before learning from any expert is likewise denoted by
\be
H_{\varnothing}=H(\varT).
\ee
To quantify how much information each expert provides to us individually we use the standard  measure for mutual information
\bea
I_{\setn{X}}&=&H_{\varnothing}-H_{\setn{X}},
\eea
in line with the basic definitional form (\ref{basic}) for information. Operationally, this equals the amount of money the expert saves by reducing their fine, by reporting their probability distribution $\Prob_X$ rather than $\Prob_{\varnothing}$.

Our experts may collaborate, as follows.
Each expert, say, Alice, Bob, and Charlie, still just measures their own variable, but 
by communicating with each other and synchronizing their measurements, they may find the joint distribution $\Prob(\varT,A,B,C)$. They can thereby jointly report $P(\varT|ABC)$. The mutual information between the joint variables and $\varT$ is then denoted by $I_{\setn{A,B,C}}$:
\bea
I_{\setn{A,B,C}}&=&H_{\varnothing}-H_{\setn{A,B,C}}.
\eea
Operationally, this equals the amount of money the three experts save themselves by collaborating.
The second term involves a single probability distribution for each set of given values $a,b,c$ of the variables $A,B,C$.
\subsection{Pooling experts' opinions}\label{poolexp}
There is another way of producing a single distribution for $\varT$ by pooling the individual experts' ``opinions'' $\Prob(\varT|x)$ for $x=a,b,c$.
One very simple (albeit inadequate) way to do this, would be to use the average distribution
\be
\Prob_{\setn{A},\setn{B},\setn{C}}(\varT)=
\frac13\left[\Prob_A(\varT)+\Prob_B(\varT)+\Prob_C(\varT)\right],
\ee
but, clearly, there are many more.
The operational idea is that we ask the three experts to report to us a single distribution, even if they never collaborated. The rule of combining different distributions $\Prob(\varT|X)$ should be symmetric between all experts, in order to conform to standard axioms required for a PID.
Note that the subscript now contains three sets, containing one expert each, reflecting that the pooled distribution makes use only of single-expert opinions $\Prob_X(\varT)$.
Once the idea of pooling is in place we can easily extend it to different types of combinations of experts. For example,
$\Prob_{\setn{A,B},\setn{C}}$ would be constructed out of the joint distribution produced by Alice and Bob collaborating and Charlie's single-expert distribution $\Prob_C$.

To make a distinction between the different types of combinations of experts it may be useful to talk about collections of sets of experts. For example, $\setn{A,B},\setn{C}$ denotes a collection of two sets, one set containing two experts, the other containing one expert.
\subsection{Synergistic, unique and redundant information}\label{SUR}
\subsubsection{Definitions}
Now let us first focus on just a pair of experts, Alice and Bob, and in what way their pooled distribution $\Prob_{\setn{A},\setn{B}}(\varT)$ determines the PID.
We first display the usual PID (sticking with standard notation for the moment) for how two variables carry information in different ways about $\varT$:
\bea \label{PID}
I_{\setn{A,B}}&=& I_{\unq{A}{B}} +I_{\unq{B}{A}} +  I_{\red{A}{B}} + I_{\syn{A}{B}},\nonumber\\
I_{\setn{A}}&=&I_{\unq{A}{B}} + I_{\red{A}{B} }\nonumber\\
I_{\setn{B}}&=&I_{\unq{B}{A}} + I_{\red{A}{B} }.
\eea
In this notation, $I_{\unq{A}{B}}$ indicates information unique to Alice w.r.t.~Bob, and $I_{\red{A}{B}}$ stands for the information redundantly encoded in (i.e., shared by)  both Alice's and Bob's probability distributions $P_{A}(\varT)$ and  $P_{B}(\varT)$, respectively.  

We define synergistic information as the difference between two entropies:
\be\label{my1}
I_{\syn{A}{B}}=H_{\setn{A},\setn{B}}-H_{\setn{A,B}},
\ee
as it captures how the whole (Alice and Bob collaborating and producing a joint distribution) is more than the sum of its parts (Alice and Bob pooling their individual distributions).
But this then determines unique information as
\bea\label{unqAB}
I_{\unq{A}{B}}=H_{\setn{B}}-H_{\setn{A},\setn{B}},
	\nonumber\\
I_{\unq{B}{A}}=H_{\setn{A}}-H_{\setn{A},\setn{B}}.	
\eea
This is again a difference between two entropies, and it gives the unique information that one expert possesses but the other does not. The operational definition is that it equals the money one expert can save for the other by pooling.

Finally, given a pooled distribution $\Prob_{\setn{A},\setn{B}}$ the redundant information is then given by
\bea\label{redun}
\Delta I_{\red{A}{B}}&=&H_{\setn{A}\setn{B}}+H_{\varnothing}
\nonumber\\
&&-H_{\setn{A}}-H_{\setn{B}}.
\eea
This is not a difference between two entropies and  it involves four (rather than two) different sorts of probability distributions over $\varT$. 
It can, of course, be written as a difference between two information quantities. This is reflected in the use of the symbol $\Delta I_{\red{A}{B}}$. This notational device has been used before, for example in Refs. \cite{schneidman2003,latham2005}, to carefully distinguish different  types of informational quantities and define synergy and redundancy in neural coding (see remark after the next Equation). 

We may also note that the above definition of redundant information is very similar to that of co-information \cite{coin} [with notation changed here by including the $\Delta$ symbol]
\bea\label{co}
\Delta I_{{\rm co}\, A\&B}&=&H_{\setn{A,B}}+H_{\varnothing}
\nonumber\\
&&-H_{\setn{A}}-H_{\setn{B}},
\eea
which features the joint distribution rather than the pooled distribution. It is well known that this quantity can take on both negative and positive values.
This quantity, in fact, has been identified with synergy and it was denoted then as $\Delta I_{{\rm syn}}$ in \cite{latham2005}.

In order to see more completely which quantities are always differences between two entropies and which one may not, 
let us now also consider an alternative way (indicated by using primed symbols) of defining the PID by first defining redundant information as a difference between two (average) entropies. We use $H_{\varnothing}$ as the higher entropy and then need an entropy $H_{\red{A}{B}}$, symmetric between A and B, that derives from some distribution that contains only information that is common to both A and B, such that
 \bea
 I'_{\red{A}{B}}=H_{\varnothing}-H_{\red{A}{B}}.
 \eea 
We do not need to specify anything further, in order to see that the unique information would then be given by
\bea
I'_{\unq{A}{B}}=H_{\red{A}{B}}-H_{\setn{B}},
\nonumber\\
I'_{\unq{B}{A}}=H_{\red{A}{B}}-H_{\setn{A}}.	
\eea 
Just as before, unique information is then a difference between two entropies. On the other hand, the synergystic information would contain four terms and we would write
\bea\label{synalt}
\Delta I_{\syn{A}{B}} &=&-H_{\setn{A,B}}-H_{\red{A}{B}}
\nonumber\\
&&+H_{\setn{A}}+H_{\setn{B}}.
\eea
In this case, then, synergy would be the odd one out, as being the only quantity in the PID that is not a difference between two entropies.  Unique information is special, in that it is truly information in either of these two case.

\subsubsection{Relation to the BROJA measure}\label{BROJ}
There are some similarities and contrasts between pooling and the way the BROJA measure---named for the authors of \cite{bertschinger2014}--- is  defined. For the BROJA measure, one first considers all joint
distributions $P_\pos(\varT,A,B)$  that are consistent with
the two marginal distributions $P(\varT,A)$ and $P(\varT,B)$. One then
maximizes over all possible joint distributions 
the entropy of $P_\pos(\varT|A,B)$. Denote that maximum  by
$\tH_{\setn{A},\setn{B}}$. The subscript here reminds us the maximum is determined by the two marginal distributions. Synergistic, unique, and redundant information are then given by
\bea
\I_{\syn{A}{B}}&=&\tH_{\setn{A},\setn{B}}-H_{\setn{A,B}},\nonumber\\
\I_{\unq{A}{B}}&=&H_{\setn{B}}-\tH_{\setn{A},\setn{B}},
\nonumber\\
\I_{\unq{B}{A}}&=&H_{\setn{A}}-\tH_{\setn{A},\setn{B}},
\nonumber\\
\Delta\I_{\red{A}{B}}&=&\tH_{\setn{A}\setn{B}}+H_{\varnothing}
\nonumber\\
&&-H_{\setn{A}}-H_{\setn{B}}.
\eea
Both the analogy and the difference between $I_{{\rm BROJA}}$ and $I_{\pool}$ are obvious. The definitions have exactly the same form as Eqs.~(\ref{my1}--\ref{redun}), but the pooling distribution is replaced by the ``worst'' [highest-entropy] possible joint distribution in all these definitions.
One consequence is that the redundant information here too is defined in terms of four different types of probability distributions, not just two.

We may also note that various inequalities derived in \cite{bertschinger2014} (see Lemma 3, in particular) do not apply to our measures, because the assumption underlying that Lemma is that such measures are derived from joint distributions consistent with the marginals.
The pooling distribution is not necessarily consistent with the marginals, as it forms a compromise instead.
For example, for the simple ``geometric average'' pooling rule mentioned below, in Eq.~(\ref{sqrtp}), averaging over variable $B$ would give us the square root of the probability distribution over $A$, renormalized.

\subsubsection{Non-negativity of information}

The original idea of the PID was to decompose mutual information into non-negative quantities, all interpretable as information.  Taking information as a reduction in uncertainty (entropy) that accompanies going from initial to final probability distributions for the target variable $\varT$, we saw that quite generically one of the four quantities introduced in (\ref{PID}) is not information. We also saw that unique information generically is indeed a difference between two entropies. As such, a requirement on unique information is that it be non-negative. [In our case, synergistic information is automatically non-negative.]

For our definition of unique information (\ref{unqAB})
the question of non-negativity boils down to the question whether Alice and Bob can pool in such a way as to ensure that their fine is not more than either Alice's or Bob's individual fine. And indeed, they can, rather trivially: they could choose to report either always Alice's distribution or always Bob's, whichever one has the lowest uncertainty.
Thus, one (easily met) requirement on the pooled distribution $\Prob_{\setn{A},\setn{B}}$ is that its expected uncertainty always be less than or at worst equal to the minimum of $H_{\setn{A}}$ and $H_{\setn{B}}$. That is, we require
\be\label{Brule}
H_{\setn{A},\setn{B}}\leq \min\left(H_{\setn{A}},H_{\setn{B}}\right).
\ee
Referring back to the very simple way of pooling by taking the average distribution, that simple method does not fulfill this criterion, as is easily checked.

In this context, given that our definition of redundant information is not, strictly speaking, information, we do not  require it to be non-negative. In fact, as we will see below in the Examples section, it can be negative for certain choices of pooling, and then it could be made positive {\em only} by increasing the uncertainty in the pooled distribution, since the other three entropies and probability distributions featuring in (\ref{redun}) are fixed. That ad-hoc fix, though, is counter to the meaning of synergy: the whole would be more than the sum of its parts, but only because we do not do our best to get as much information as we can from the parts.

For two source variables $A$ and $B$ it may well be that the intuitions behind unique, redundant and synergistic information (as defined within PID) are  incompatible. That is, whenever one defines one of the three quantities, the remaining two are fixed, but that particular way of fixing their magnitudes may not be in agreement with what the remaining two terms are supposed to mean. The idea that not all quantities are expressible as a difference between two entropies may be taken as an indication in that direction as well.
\subsection{Pooling-based lattice}\label{lattice}
The original work on PID \cite{williams2010} argued for a particular lattice underlying the PID structure. That, is, the concept of redundant information naturally leads to a partial order, as well as that of a unique least upper bound and a unique greatest lower bound that can be assigned to every pair of elements. One idea behind the construction of a lattice, formulated in terms of our experts from Section II, is as follows: if we have a collection of sets of experts, and within that collection, one set is a subset of another, the redundant information is already present in the subset, and so we can delete the superset. The lattice of collections  that thus remains is the redundancy-based lattice. 

Here we use pooling instead of redundancy as our fundamental notion, and this leads to the opposite approach: 
if within a collection of sets of experts, one set is a subset of another, then we should use the superset when we pool, not the subset. So we delete the subset from our collection. The collections remaining form a pooling-based lattice.

For example, for three variables (or experts) neither lattice contains the collection $\setn{A,B}, \setn{A,C}, \setn{B}$.
In the redundancy-based lattice the set $\setn{A,B}$ is removed; in the pooling-based lattice, the set $\setn{B}$ is removed. 

For two elements in the pooling-based lattice, for example, $\setn{A}$ and $\setn{B,C}$, one defines the least upper bound and the greatest lower bound as follows.
The least upper bound is the collection $\setn{A},\setn{B,C}$, which corresponds to pooling the probability distributions from Alice with that of Bob and Charlie jointly.
The information thus obtained is larger or equal to that of the maximum obtainable from either Alice, or by Bob and Charlie jointly. 
The greatest lower bound is the empty collection $\varnothing$. The information obtainable from Alice, and from Bob and Charlie jointly, are both equal to or more than that obtainable from not consulting any expert. There is no larger collection of experts with that property.

With every node of the lattice we associate probability distributions  (over $\varT$, one distribution for each of the values the other variables in the collection may take) and a number, namely, their average entropy.
The empty set is the least element in the pooling-based lattice, and we associate
the distribution $\Prob_{\varnothing}(\varT)$ and its entropy $H_{\varnothing}$ with it. 
Refs.~\cite{ay2019,rosas2020} derive the same lattice from different considerations.

\section{Pooling probability distributions}\label{pool}
For two variables the PID depends on the probability distributions $P_{\setn{a,b}}(\varT)$ two experts Alice and Bob agree to use if all they know are the marginal distributions and the corresponding conditional distributions $P_A(\varT)=P(\varT|A)$ and
$P_B(\varT)=P(\varT|B)$. 

There are several options, some involving optimizations, others simpler.  In most examples a simple geometric averaging procedure is used to define the pooled distribution. When that procedure fails a more general one-parameter family of pooled distributions is used and optimized over that one parameter.
\subsection{Minimum goal}\label{minimum}

We consider here a simple example where some previous measures of synergy disagree. But let us first see what the essence is of an even simpler example where there is consensus on synergistic information. If $A$ and $B$ are random bits and $\varT=A\oplus B$ then everyone agrees there is only one bit of synergistic information present.
The reason is simple, neither $\Prob_A(\varT)$ nor $\Prob_B(\varT)$, which are identical, contains any information, and only the joint distribution yields the full one bit of information through
$\Prob(\varT|A,B)$.

In contrast, consider the example from Table  \ref{tabel1}. The two variables $A,B$ are binary, the target variable is ternary, and there are only three possible combinations of values with nonzero probability.
\begin{table}
	\begin{center}
		\begin{tabular}{ |c|c| } 
			\hline
			$A,B,\varT$ &		Probability  \\ 
			\hline
			$0,0,0$&		1/3  \\ 
			$0,1,1$&		1/3  \\ 
			$1,0,2$&		1/3 \\ 
			\hline
		\end{tabular}
	\end{center}

		\begin{center}
		\begin{tabular}{ |c|c|c|c|c|} 
			\hline
			
			$I_\partial$ &$I_{{\rm BROJA}}$ & $I_{\ccs}$& $I_{\min}$	& $I_{\pool}$  \\ 
			\hline
			$\syn{A}{B}$ &	$0$ &	$0.138$ &	$0.333$   & $0$\\ 
			$\unq{A}{B}$&	$0.667$&	0.528 & $0.333$  & $0.667$\\ 
			$\unq{B}{A}$ &	$0.667$&	0.528  & $0.333$  & $0.667$\\ 
			$\red{A}{B}$ & $0.252$& 0.390  & $0.585$  & $0.252$\\
			\hline
		\end{tabular}
	\end{center}
		\caption{Simple example (5A from \cite{ince2017}) in which the marginal distributions $\Prob(\varT,A)$ and $\Prob(\varT,B)$  together contain as much information about the target variable as does the joint distribution $\Prob(\varT,A,B)$ .  
					In this case, the BROJA measure agrees exactly (but via a different mechanism, see Section \ref{BROJ}) with the pooling measure. In particular,  there is no synergistic information.}\label{tabel1}
\end{table}
The interesting (debatable) situation occurs when both $A$ and $B$ have the value 0.
We are certain in this case of the value of $\varT$, even without knowing the joint distribution. This is because each variable eliminates one possibility, leaving just one value of $\varT$ to occur with 100\% probability.
There is no synergistic information here, since the joint distribution $P(A,B,\varT)$ cannot give us more information than we can obtain from the marginals $P(A,\varT)$  and $P(B,\varT)$ .

Some measures (in particular, from those measures that we compare to later on in Section \ref{Examples}, we find that $I_{\min}$ from \cite{williams2010} and $I_{\ccs}$ from 
\cite{ince2017}) ascribe a nonzero amount of synergistic information to this case. On the other hand, the BROJA measure (mentioned above) does yield zero synergistic information.

What the simple example means for our pooling rules is two-fold:
\begin{enumerate}
	\item If for some combination of the values of the variables all marginal distributions are identical, then the pooled distribution for that combination will have to equal that distribution.
	\item If for a particular combination of variables one of the  marginal distributions indicates a certain value of $\varT$ appears with 0\% probability, then this should be true for the pooled distribution as well.
\end{enumerate}
These are taken as two necessary (although by no means sufficient) requirements on pooling, in addition to the requirement (\ref{Brule}) we found above.

\subsection{Simple pooling rules}\label{simple}
There are a few different rules that have been proposed for pooling expert opinions, with different circumstances  leading to different conclusions about which way is appropriate. (See Refs.~\cite{lind1988,pool2000,carvalho2023,neyman2022} for background information about this topic.)

Here are two ways of pooling that seem inapplicable to our specific case, where each expert opinion $\Prob(\varT|X)$ for $X=A,B,C,\ldots$ is derivable from a joint distribution $\Prob(\varT,A,B,C,\ldots)$.
First, there is the (weighted) average
\be
P(\varT|A,B,C,\ldots)=\sum_X w_X \Prob(\varT|X), 
\ee
with $w_X>0$ and $\sum_X w_X=1$.
This may be correct when experts may contradict each other and a compromise is needed. But in our case it fails the second rule.

Another rule would be to take the product
\be
P(\varT|A,B,C,\ldots)=\Pi_X \Prob(\varT|X).
\ee
This would apply to a case where we try to estimate the average value of $\varT$ and we have independent data that, nonetheless, do estimate the same average. Then it is true that the more data we have the better our average should be determined. The variance in the distribution of possible values for the average should decrease with more data. But this rule violates our first requirement on pooling.
\subsubsection{Geometric average}
The third rule does fit both our pooling requirements and, moreover, has other advantages, which, however, do not concern us here \cite{lind1988,pool2000,carvalho2023,neyman2022}.
Although the following definition used here can easily be extended to more than two probability distributions (or agents) we focus here on the case of two agents, Alice and Bob, first. (See Appendix A about the extension to more than two experts.)
We define for each pair of values $a,b$ for $A,B$ the geometric average
\be\label{sqrtp}
P_{\setn{A}\setn{B}}(\varT|a,b)=\sqrt{\Prob(\varT|a)\Prob(\varT|b)}/Z_{ab},
\ee
where  the normalization factor
\be\label{ZZ}
Z_{ab}=\sum_t \sqrt{\Prob(t|a)\Prob(t|b)},
\ee
is needed to define a proper probability distribution.
$Z_{ab}$ is an overlap between two distribution functions, and is in fact the \bhat measure \cite{bhat1946}.
It lies between 0 (for orthogonal distributions, which have no common support) and 1, for identical distributions.

The expected value of Alice's and Bob's  fine  is
\be
H_{\setn{A}\setn{B}}=\frac12 H_{\setn{A}} +\frac12 H_{\setn{B}}-\B,
\ee
with the non-negative quantity $\B$ given by
\be
\B=\sum_{a}\sum_{b} \Prob(a,b)\log(1/Z_{ab}),
\ee
where 
\be
\Prob(a,b)=\sum_t \Prob(t,a,b)
\ee
is the joint distribution for $a,b$.
Note that an average fine equal to the sum of the first two terms can be obtained by the agents simply by reporting either $P_A$ or $P_B$ with 50\% probability. That is why we may consider $\B$ as a ``bonus,'' extra money saved by Alice and Bob when they pool their probability distributions instead of randomly choosing $P_A$ or $P_B$. 

Note that Alice and Bob cannot determine the value of the bonus $\B$ as long as they do not know the joint distribution $\Prob(a,b)$. They could still determine the {\em minimum} bonus by minimizing over all possible joint distributions $\Prob_\pos(a,b)$ consistent with $\Prob(a)$ and $\Prob(b)$. That is one way for them to see if the bonus is large enough to satisfy the  inequality (\ref{Brule}), i.e., whether
\be \label{BB}
\B\geq \frac12 \left|H_{\setn{A}}-H_{\setn{B}}\right|.
\ee
If one needs to define a PID for a system for which one actually knows all probability distributions, one could be satisfied with requiring (\ref{BB}) to hold for the actual distribution $\Prob(a,b)$.

Now that we have an expression for $H_{\setn{A}\setn{B}}$ we can write the following relation between unique information and the bonus,
\be
I_{\unq{A}{B}}+I_{\unq{B}{A}}=
2\B,
\ee
which neatly quantifies the intuition that unique information is determined by the extent to which the probability distributions $\Prob_{\setn{A}}$ and $\Prob_{\setn{B}}$ differ.

The individual unique information quantities are given by
\bea
I_{\unq{A}{B}}&=& \B +\frac12\left( H_{\setn{B}}-H_{\setn{A}} \right),\nonumber\\
I_{\unq{B}{A}}&=& \B +\frac12\left( H_{\setn{A}}-H_{\setn{B}}\right),
\eea
which are non-negative thanks to the requirement (\ref{BB}).
\subsubsection{One-parameter family}
This method of pooling is called ``logarithmic'' and generalizes easily to using pooling distributions of the form
\be\label{weight}
P^{(w)}_{\setn{A}\setn{B}}(\varT|a,b)=\Prob(\varT|a)^w\Prob(\varT|b)^{1-w}/Z_{w,ab},
\ee
for $0\leq w \leq 1$, with normalization
\be
Z_{w,ab}=\sum_t \Prob(t|a)^w\Prob(t|b)^{1-w}.
\ee
The expected fine can be written as
\be
H_{\setn{A}\setn{B}}=wH_{\setn{A}} +(1-w) H_{\setn{B}}-\B_w,
\ee
with 
\be
\B_w=\sum_{a}\sum_{b} \Prob(a,b)\log(1/Z_{w,ab}),
\ee
which is non-negative, as can be shown using the  Rogers-H\"older's inequality \cite{maligranda1998}. (For the simpler case of $w=1/2$ adopted above, this inequality reduces to the more straightforward  Cauchy-Schwarz inequality.)
 
In general, we recommend using the one-parameter family (\ref{weight}) of pooling distributions if the simple geometric average ($w=1/2$) violates condition (\ref{Brule}). One way to choose the weight $w$ would be to
make it dependent on the entropies $H_{\setn{A}}$ and $H_{\setn{B}}$. For example, one might choose
$w=2^{-H_{\setn{A}}}/(2^{-H_{\setn{A}}}+2^{-H_{\setn{B}}})$, thus giving more weight to the lower-entropy distribution. Another way is discussed next. It is important to note that there is always a value of $w$ such that all our requirements on the pooling distribution are met, namely, those listed at the end of subsection \ref{minimum} and (\ref{Brule}).  
\subsection{Optimized pooling}
Denote the set of possible joint distributions $P_\pos(\varT,A,B)$ consistent with $P_A$ and $P_B$ as $\Delta_P$.
Denote the set of all joint probability distributions by $\Omega$. 
That is, $\Delta_P\subset\Omega$.
If Alice and Bob choose a distribution $\omega(\varT,A,B)$ from $\Omega$, then, relative to a possible distribution
$P_\pos(\varT,A,B)$ their expected fine is given by
\bea
F(\omega,\Prob_\pos):=-\sum_a \sum_b \sum_t P_\pos(t,a,b) \log \frac{\omega(t,a,b)}{\omega(a,b)}\nonumber
\eea
with $\omega(a,b)=\sum_t \omega(t,a,b)$.
They would like to minimize their fine, although they do not know which possible distribution $\Prob_\pos \in \Delta_P$ they have. They could  find the worst-case distribution $\Prob_\pos$ for each choice of $\omega$, and then minimize over all possible choices of $\omega$.
Their ``best'' fine would then be
\bea
F_\best=\min_{\omega\in\Omega} \max_{\Prob_\pos\in\Delta_P}
F(\omega,\Prob_\pos).
\eea
A simpler version of this idea would make use of a smaller subset of all distributions $\omega$.
For example, they may restrict to the type of logarithmic pooling distributions discussed above, but with arbitrary weights;
\be
\frac{\omega(t,a,b)}{\omega(a,b)}=\Prob(t|a)^{w}
\Prob(t|b)^{1-w},
\ee
for all $0\leq w\leq 1$, and then perform the minimization just over the parameter $w$. In the next Section with examples, we will display one case where $w=1/2$ violates the requirement (\ref{BB}). That is, the pooling distribution $\sqrt{\Prob(\varT|a)\Prob(\varT|b)}/Z_{ab}$
is inferior to the better one of  $\Prob_{\setn{A}}$ and  $\Prob_{\setn{B}}$. But minimization over $w$ leads to a pooling distribution of the form (\ref{weight}) superior to both  $\Prob_{\setn{A}}$ and  $\Prob_{\setn{B}}$, which does satisfy all our requirements. 
 
\section{Examples}\label{Examples}
In the following $I_{\pool}$ is based on logarithmic pooling with weight $w=1/2$, unless stated otherwise.
\subsection{Standard examples}

\begin{table}[h]\label{AND}
	\begin{center}
		\begin{tabular}{ |c|c| } 
			\hline
			$A,B,\varT$ &		Probability  \\ 
			\hline
			$0,0,0$&		1/4  \\ 
			$0,1,0$&		1/4  \\ 
			$1,0,0$&		1/4 \\ 
			$1,1,1$& 1/4\\
			\hline
		\end{tabular}
	\end{center}

	\begin{center}
		\begin{tabular}{ |c|c|c|c|c|} 
			\hline
			
			$I_\partial$ &$I_{{\rm BROJA}}$ & $I_{\ccs}$& $I_{\dep}$	& $I_{\pool}$  \\ 
			\hline
			$\syn{A}{B}$ &	$0.500$ &	$0.292$ &	$0.270$   & $0.250$\\ 
			$\unq{A}{B}$&	$0$&	0.208  & $0.230$  & $0.250$\\ 
			$\unq{B}{A}$ &	$0$&	0.208  & $0.230$  & $0.250$\\ 
			$\red{A}{B}$ & $0.311$& 0.104  & $0.082$  & $0.061$\\
			\hline
		\end{tabular}
	\end{center}
	\caption{Comparison of a few PIDs for the AND example. $I_\pool$ is close to $I_\dep$ and to a lesser degree to $I_\ccs$, but clearly disagrees with the BROJA measure. See Table 5 from Ref.~\cite{james2018b}}.\label{PIDs}
\end{table}

Many examples of distributions have been tested on many different PIDs. Here we just use a small selection of three examples, on a small selection of PIDs to compare the PID that results from the simplest pooling strategy.
The results for other measures used here are all conveniently found in Table 5 from Ref.~\cite{james2018b}.
One point of that Table was to show how various measures clearly disagree with the original measure $I_{\min}$ from \cite{williams2010}, and it turns out the same is true for $I_\pool$.
For the TWO BIT example $I_\pool$ agrees with all three measures, and so is not displayed here. The comparisons for the NOT TWO example are similar to those for AND, and are for that reason not displayed here either.

The three examples we do consider are in Tables II, III, and IV.  One point here is to show that there is always an example where $I_{\pool}$ disagrees with a given measure. On the other hand, for those same examples there also are other measures $I_{\pool}$ agrees with, either exactly or approximately.

\begin{table}[h]\label{DIFF}
	\begin{center}
		\begin{tabular}{ |c|c| } 
			\hline
			$A,B,\varT$ &		Probability  \\ 
			\hline
			$0,0,0$&		1/4  \\ 
			$0,0,1$&		1/4  \\ 
			$0,1,0$&		1/4 \\ 
			$1,0,1$& 1/4\\
			\hline
		\end{tabular}
	\end{center}

	\begin{center}
		\begin{tabular}{ |c|c|c|c|c|} 
			\hline
			
			$I_\partial$ &$I_{{\rm BROJA}}$ & $I_{\ccs}$& $I_{\dep}$	& $I_{\pool}$  \\ 
			\hline
			$\syn{A}{B}$ &	$0$ &	$0.085$ &	$0$   & $0$\\ 
			$\unq{A}{B}$&	$0.189$&	0.104  & $0.189$  & $0.189$\\ 
			$\unq{B}{A}$ &	$0.189$&	0.104  & $0.189$  & $0.189$\\ 
			$\red{A}{B}$ & $0.123$& 0.208  & $0.123$  & $0.123$\\
			\hline
		\end{tabular}
	\end{center}
	\caption{Comparison of a few PIDs for the DIFF example. $I_\pool$ agrees exactly with $I_\dep$ and  $I_{{\rm BROJA}}$, but disagrees with $I_{\ccs}$. See Table 5 from Ref.~\cite{james2018b}.}
\end{table}

\begin{table}[h]\label{PU}
	\begin{center}
		\begin{tabular}{ |c|c| } 
			\hline
			$A,B,\varT$ &		Probability  \\ 
			\hline
			$0,1,1$&		1/4  \\ 
			$1,0,1$&		1/4  \\ 
			$0,2,2$&		1/4 \\ 
			$2,0,2$& 1/4\\
			\hline
		\end{tabular}
	\end{center}

	\begin{center}
		\begin{tabular}{ |c|c|c|c|c|} 
			\hline
						$I_\partial$ &$I_{{\rm BROJA}}$ & $I_{\ccs}$& $I_{\dep}$	& $I_{\pool}$  \\ 
			\hline
			$\syn{A}{B}$ &	$0.500$ &	$0$ &	$0.250$   & $0$\\ 
			$\unq{A}{B}$&	$0$&	0.500  & $0.250$  & $0.500$\\ 
			$\unq{B}{A}$ &	$0$&	0.500  & $0.250$  & $0.500$\\ 
			$\red{A}{B}$ & $0.500$& 0  & $0.250$  & $0$\\
			\hline
		\end{tabular}
	\end{center}
	\caption{Comparison of a few PIDs for the PNT.~UNQ. example. $I_\pool$ agrees with $I_\ccs$. See Table 5 from Ref.~\cite{james2018b}.}\label{PIDs}
\end{table}

\subsection{When the geometric average fails}
For the example of Table V (found numerically, then simplified to make all percentages integers) the pooling distribution obtained by taking the geometric average of Alice's and Bob's individual distributions fails to meet requirement (\ref{BB}) or, equivalently, (\ref{Brule}).

That is, Alice and Bob would be better off (i.e., incur a smaller fine on average) simply always reporting the more informative individual distribution, in this case Bob's (with an average fine (entropy) of 0.321 bits vs 0.472 bits for Alice's). They would be aware of the flaw in the simple pooling method, and so could indeed report Bob's distribution $P_{\setn{B}}(\varT)$. In that case, the unique information from Alice would be identically zero (as she does not contribute anything to the pooling distribution).
	
	 However, by considering the logarithmic pooling rule with arbitrary unequal weights, they would find that by using a weight $w\approx 0.12$ for Alice and hence a weight $1-w\approx 0.88$ for Bob, they would do even better, as displayed in the Table V. Alice would thus contribute a very small but nonzero amount of unique information. 
\begin{table}[h]
	\begin{center}
	\begin{tabular}{ |c|c| } 
		\hline
		$A,B,\varT$ &		Probability  \\ 
		\hline
		$0,0,0$&		0.50  \\ 
		$1,0,0$&		0.39  \\ 
		$1,1,0$&		0.01 \\ 
		$0,0,1$& 0.04\\
				$0,1,1$& 0.04\\
						$1,0,1$& 0.01\\
								$1,1,1$& 0.01\\
		\hline
	\end{tabular}
\end{center}

	\begin{center}
	\begin{tabular}{ |c|c|c|} 
		\hline
		
		$I_\partial$ &$I_{\pool}$, $w=1/2$ & $I_{\pool}$, $w=0.12$  \\ 
		\hline
		$\syn{A}{B}$ &	$0.051$ &	$0.026$ \\ 
		$\unq{A}{B}$&	$-0.023$&	0.002  \\ 
		$\unq{B}{A}$ &	$0.108$&	0.133   \\ 
		$\red{A}{B}$ & $0.040$& 0.015  \\
		\hline
	\end{tabular}
\end{center}
	\caption{Comparison of two different logarithmic pooling distributions.}
\end{table}\label{fail}

\subsection{Negative $\Delta I_{\red{A}{B}}$}
Table VI presents an example where the simple pooling method leads to negative redundant information, while still meeting requirement (\ref{BB}) or, equivalently, (\ref{Brule}). As mentioned before, this can be seen as a consequence of that quantity depending on four different probability distributions, which means its interpretation as ``information'' is not straightforward.
The example is called ReducedOr in \cite{ince2017} where it is attributed to Joseph Lizier. The point of that example was to locate a defect in the BROJA measure. That is, the example clearly contains unique information from both variables, even though the BROJA measure (as well as the original $I_{\min}$ measure from Ref.~\cite{williams2010}) assigns zero unique information.
In fact, when either variable takes on the value 1,  it provides the unique information that the target variable must have the value 1.

Our pooling measure agrees with that verdict, but the main reason for displaying this example here, is that the shared (or redundant) information is negative. Recalling the expression for redundant information, we could make it equal to zero only in an ad-hoc manner, by increasing the entropy of the pooled distribution by 0.09 units (and then all 4 quantities would numerically agree with $I_{\ccs}$, as is easily checked.) Since increasing that uncertainty would correspond to Alice and Bob increasing their fine, this would go against the spirit of pooling. In any case, as noted several times, the fact that $\Delta I_{\red{A}{B}}$ is not simply a difference between two entropies, and is thus not information, also (arguably) eliminates the requirement for it to be non-negative.
We thus accept here the result displayed in Table VI.
\begin{table}[h]
	\begin{center}
		\begin{tabular}{ |c|c| } 
			\hline
			$A,B,\varT$ &		Probability  \\ 
			\hline
			$0,0,0$&		1/2  \\ 
			$1,0,1$&		1/4  \\ 
			$0,1,1$&		1/4 \\ 
			\hline
		\end{tabular}
	\end{center}
	\begin{center}
	\begin{tabular}{ |c|c|c|c|} 
		\hline
		
		$I_\partial$ &$I_{{\rm BROJA}}$ & $I_{\ccs}$& $I_{\pool}$  \\ 
		\hline
		$\syn{A}{B}$ &	$0.69$ &	$0.38$ & $0.29$\\ 
		$\unq{A}{B}$&	$0$&	0.31  & $0.40$\\ 
		$\unq{B}{A}$ &	$0$&	0.31   & $0.40$\\ 
		$\red{A}{B}$ & $0.31$& 0  & $-0.09$\\
		\hline
	\end{tabular}
\end{center}
\caption{Comparison of a few PIDs for the ReducedOr example discussed in \cite{ince2017}. $I_\pool$ agrees with $I_\ccs$ against $I_{{\rm BROJA}}$ that there is nonzero unique information in this case.).}\label{PIDs}
\end{table}\label{RedOr}

\section{Conclusions}
Within the context of the Partial Information Decomposition (PID, \cite{williams2010}) synergistic information is meant to  quantify how much ``the whole'' is more than ``the sum of its parts.'' In the case of two source variables $A$ and $B$ that provide information about a target variable $\varT$, we identified here ``the parts'' with marginal probability distributions $\Prob(A,\varT)$ and $\Prob(B,\varT)$ and ``the whole'' with the joint probability distribution $\Prob(A,B,\varT)$. 
Synergistic information is then the average reduction in uncertainty (as measured by entropy) upon using the  joint distribution rather than a specific distribution aggregated (or pooled) from the two marginal distributions for making predictions about $\varT$.

This idea of pooling \cite{lind1988} was shown to lead to a lattice underlying the PID, which differs from the original redundancy-based lattice used by Williams and Beer in their original work proposing the PID \cite{williams2010}. Each element of the lattice is a collection of sets of the variables, with no set in the collection being a subset of another. With each element of the lattice we can associate pooled probability distributions, their average entropy, and synergistic information. For example, given the collection consisting of $\setn{A,B}$ and $\setn{C}$, we combine
$\Prob(\varT|A,B)$ and $\Prob(\varT|C)$ into pooled distributions, and calculate their average entropy (averaged over all variables $A,B,C,\varT$).
The difference between the average entropy of the pooled distributions and the entropy of the full  distribution $\Prob(\varT|A,B,C)$ is defined to equal the synergistic information associated with that element of the lattice.

We considered logarithmic pooling as a simple and convenient pooling method which can be optimized to provide the ``best'' logarithmic way to pool information. It provides us with sensible  definitions of synergistic information as well as of unique information, both satisfying the basic definition of information as a reduction in uncertainty when switching from one type of probability distribution to a (better) one.
Inevitably, ``redundant information'' then involves four different types of probability distributions and thus is not of that basic form.  The possible pooling distributions (\ref{weight}) are parametrized by just a single parameter $w$, and the only requirement on $w$ is that the condition (\ref{Brule}) be fulfilled. That condition is equivalent to requiring unique information to be non-negative. There is always such a value for $w$.

Other (more complicated) ways of pooling distributions are possible, and depending on one's definition of the ``best'' way to do this, one finds different measures of synergistic information and hence of unique information. This freedom of choice  illustrates  the ambiguity in the definition of a PID.

\section*{Appendix A: More than two sources}\label{more}
The idea of logarithmic pooling is easily extended to more than two experts, as follows. For experts $A,B,C,\ldots$ we may define
\be
\Prob_{\setn{A},\setn{B},\setn{C},\ldots}(\varT)
=\Prob_A(\varT)^{w_A}
\Prob_B(\varT)^{w_B}
\Prob_C(\varT)^{w_C}\ldots
\ee
where the positive weights $w_A,w_B,w_C,\ldots$ add up to unity. For three experts one may wonder whether to choose the weights equal, as we did for two experts by default. 
There is a good reason not to always do this: once we have calculated the redundant information shared amongst pairs of experts, we may then penalize such redundancy by lowering the sum of the weights for such pairs.
Thus, if Alice and Bob's information is purely shared, and Charlie's information is unique relative to Alice and Bob, then it makes sense to set $w_C=1/2$ and $w_A=w_B=1/4$, for example. That is, Alice and Bob together are assigned the same weight as Charlie.
Since this case provides one extreme, we may set the general rule for three experts:
\bea\label{wbounds}
\frac14 \leq w_k \leq \frac12
\eea
where within this range there is freedom depending on how one takes into account the pairwise redundant and unique information.

For more than three experts (say, $N_e$ of them) it becomes more complicated to design general rules for assigning weights. The upper bound of 1/2 still holds (and would be correct only if all experts but one provide only shared information: that one expert gets the weight 1/2; the other $N_e$-1 experts share equally the remaining weight) and we may thus generalize (\ref{wbounds}) as
\be
\frac{1}{2(N_e-1)}\leq w_k \leq \frac{1}{2}.
\ee

The idea of pooling for more than two experts still leads to a lattice, as explained before in Section \ref{lattice}.

\bibliographystyle{unsrt} 
\bibliography{PID}

\end{document}